\documentclass{article}[11pt]
\usepackage{mathpazo}

\usepackage{mathtools}
\usepackage{amssymb}
\usepackage{amsmath}
\usepackage{amscd}
\usepackage{amsthm}
\usepackage{bm}
\usepackage{tikz}

\let\oldmarginpar\marginpar
\renewcommand\marginpar[1]{\-\oldmarginpar[\raggedleft\footnotesize #1]{\raggedright\footnotesize #1}}

\title{On the Hessian of Shape Matching Energy}
\author{Yun Fei}
\date{}
\begin{document}

\maketitle

\begin{figure}
\includegraphics[width=\linewidth]{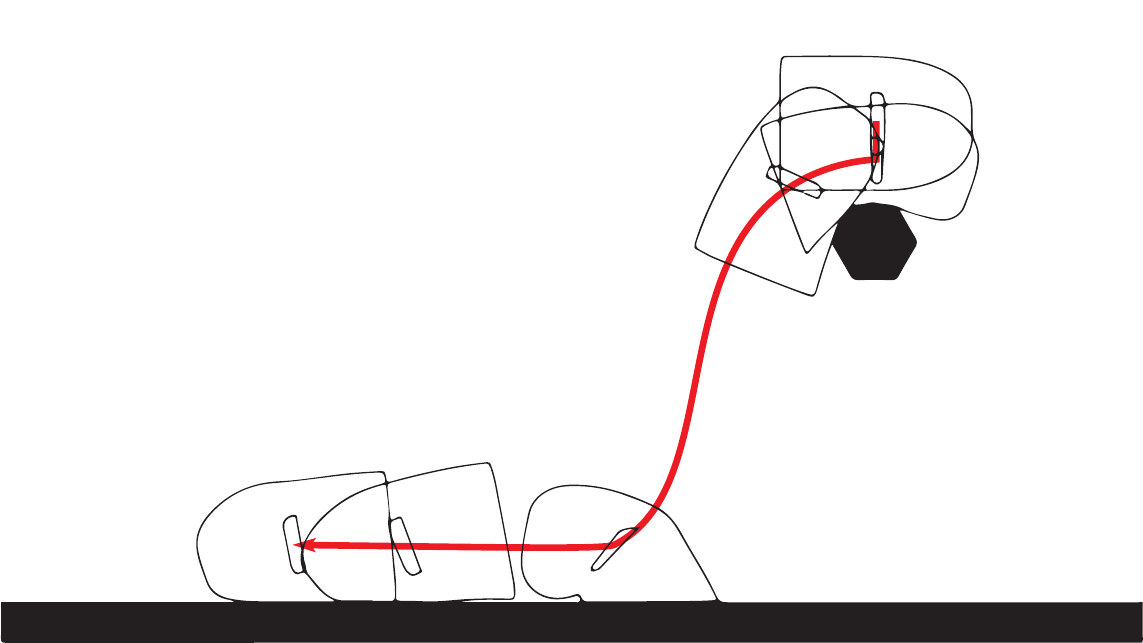}
\caption{Deformation done with implicit shape matching.}
\label{fig:sm} 
\end{figure}

\section{Introduction}
In this technical report we derive the analytic form of the Hessian matrix for shape matching energy. Shape matching (Fig.~\ref{fig:sm}) is a useful technique for meshless deformation, which can be easily combined with multiple techniques in real-time dynamics (refer to~\cite{Muller2005,bender2015position} for more details). Nevertheless, it has been rarely applied in scenarios where implicit (such as backward differentiation formulas) integrators are required, and hence strong viscous damping effect, though popular in simulation systems nowadays, is forbidden for shape matching. The reason lies in the difficulty to derive the Hessian matrix of the shape matching energy. Computing the Hessian matrix correctly, and stably, is the key to more broadly application of shape matching in implicitly-integrated systems.

\section{Shape Matching Energy for Point Cloud}
Given the world-space positions of the $r$-th vertex on the object as $\mathbf{q}_r$, and its local coordinate of the rest pose as $\mathbf{q}_r^0$, we define the shape matching potential as 
\begin{equation}
V=\frac{1}{2}\sum_rk_r\mathbf{d}_r^T\mathbf{d}_r
\end{equation}
and
\begin{equation}
\mathbf{d}_r=\mathbf{q}_r-B(\mathbf{q}_r)\mathbf{q}_r^0-t
\end{equation}
where
\begin{equation}
t=\frac{1}{M}\sum_rm_r\mathbf{q}_r
\end{equation}
is the center of mass and 
\begin{equation}
B(\mathbf{q}_r)=\gamma A_a(\mathbf{q}_r)A_s^{-1}+(1-\gamma)[R, 0](\mathbf{q}_r)
\end{equation}
is the blending between the covariance matrix $A$ and the best-fit rotation matrix $R$. $A_a$ is the asymmetric part where
\begin{equation}
A_a=\frac{1}{M}\sum_rm_r(\mathbf{q}_r-t)(\mathbf{q}_r^0)^T
\end{equation}
and 
\begin{equation}
A_s=\frac{1}{M}\sum_rm_r\mathbf{q}_r^0(\mathbf{q}_r^0)^T
\end{equation}

Usual methods extracting the rotation include using the singular value decomposition (SVD), polar decomposition or QR decomposition to factorize the covariance matrix. The choice between these methods have been extensively discussed in previous literature, where the polar decomposition has been proven to be numerically stable against small perturbation~\cite{shoemake1992}, which fits our need to compute the Jacobian and Hessian of the rotation matrix. Given the asymmetric part of a covariance matrix, $A_a$, its polar decomposition can be written as
\begin{equation}
\label{equ:ars}
A_a(\mathbf{q})=R(\mathbf{q})S(\mathbf{q})
\end{equation}
where $R$ is the rotational part.

Taking the derivative of the potential energy over $\mathbf{q}_i$, we get the potential gradient to be applied on vertex $i$, which has the following form:
\begin{equation}
\label{equ:nablav}
\nabla_i V=\sum_rk_r(\frac{\partial\mathbf{d}_r}{\partial\mathbf{q}_i})^T\mathbf{d}_r
\end{equation}
where
\begin{equation}
\frac{\partial\mathbf{d}_r}{\partial\mathbf{q}_i}=
\begin{cases}
I-\frac{\partial B\mathbf{q}_r^0}{\partial \mathbf{q}_i}-\frac{m_i}{M}I & i=r \\ 
-\frac{\partial B\mathbf{q}_r^0}{\partial \mathbf{q}_i}-\frac{m_i}{M}I & i\neq r 
\end{cases}
\end{equation}
and
\begin{equation}
\frac{\partial B\mathbf{q}_r^0}{\partial \mathbf{q}_i}=\gamma\frac{\partial A_a}{\partial \mathbf{q}_i}A_s^{-1}\mathbf{q}_r^0+(1-\gamma)[\frac{\partial R}{\partial \mathbf{q}_i}, 0]\mathbf{q}_r^0
\end{equation}
while
\begin{equation}
\frac{\partial A_a}{\partial \mathbf{q}_i}=\frac{m_i}{M}(\mathbf{q}_i^0-\sum_r\frac{m_r}{M}\mathbf{q}_r^0)^T\otimes I
\end{equation}
where $\otimes$ is the tensor product.

One naively using symbolic differentiation to solve for Jacobian of the rotation $\frac{\partial R^T\mathbf{q}_r^0}{\partial \mathbf{q}_i}$ as well as the Hessian will sooner or later meet lots of numerical singularities, even for some very simple cases. In literatures such as~\cite{mathai1997jacobians,papadopoulo2000estimating,twigg2010point} the Jacobian of SVD is derived but not for its Hessian; besides several degenerated cases need to be specifically handled. In~\cite{barbivc2011real} the authors derived the Hessian of polar decomposition for a single temporal derivative. Here we generalize their results to partial derivatives, over positions, which are also specifically simplified to avoid tensor algebra.

Multiplying both sides of Equ.~\ref{equ:ars} with $R^T(\mathbf{p})$ fixed at some point $\mathbf{p}$, we have
\begin{equation}
\label{equ:rtrs}
R^T(\mathbf{p})A_a^*(\mathbf{q})=(R^T(\mathbf{p})R(\mathbf{q}))S(\mathbf{q})
\end{equation}
where $A_a^*(\mathbf{q})$ is the first two columns (in 2D, or three columns in 3D) of $A_a$ (the zero-order term). Since $R^T(\mathbf{p})R(\mathbf{q})$ is identity for $\mathbf{p}=\mathbf{q}$, its derivative at $\mathbf{p}=\mathbf{q}$ must be some skew-symmetric matrix $\tilde{\omega}$ for some vector $\omega$. We have
\begin{equation}
\frac{\partial R(\mathbf{q})}{\partial \mathbf{q}}=R(\mathbf{q})\tilde{\omega}
\end{equation}
where we define $\tilde{\omega}x=\omega\times x$ and $\text{skew}(A)$ as the unique vector $\omega$ such that $\tilde{\omega}=(A-A^T)/2$. Then by differentiating both side of Equ.~\ref{equ:rtrs}, we have
\begin{equation}
R^T(\mathbf{p})\frac{\partial A_a^*(\mathbf{q})}{\partial\mathbf{q}}=\tilde{\omega}S(\mathbf{p})+\frac{\partial S(\mathbf{q})}{\partial\mathbf{q}}
\end{equation}
Now replacing $\mathbf{p}$ with $\mathbf{q}$, and applying the skew operator to both sides, we drop the symmetric term $\frac{\partial S(\mathbf{q})}{\partial\mathbf{q}}$, and we have
\begin{equation}
\frac{1}{2}(tr(S)I-S)R^T\omega=\text{skew}(R^T\frac{\partial A_a^*}{\partial \mathbf{q}})
\end{equation}
where we define
\begin{equation}
G=(tr(S)I-S)R^T,
\end{equation}
and equation
\begin{equation}
G\omega_{ij}=2\text{skew}(R^T\frac{\partial A_a^*}{\partial \mathbf{q}_{ij}})
\end{equation}
can be solved for $\omega_{ij}$. We also define $\mathbf{e}_j$ as the vector where the $j$-th element is $1$ and all zeros for other elements, we have
\begin{equation}
\frac{\partial A_a^*}{\partial \mathbf{q}_{ij}}=\frac{m_i}{M}\mathbf{e}_j(\mathbf{q}_i^0)^T
\end{equation}n 2D) of 
therefore
\begin{equation}
\label{equ:Gomega}
G\omega_{ij}=\frac{2m_i}{M}\text{skew}(R^T\mathbf{e}_j(\mathbf{q}_i^0)^T)
\end{equation}
Solving for $\omega_{ij}$, then we can compute
\begin{equation}
\frac{\partial R}{\partial \mathbf{q}_{ij}}=\tilde{\omega}_{ij}R
\end{equation}
and correspondingly for 3D case
\begin{equation}
\frac{\partial R\mathbf{q}_r^0}{\partial \mathbf{q}_{i}}=[\tilde{\omega}_{i0}R\mathbf{q}_r^0, \tilde{\omega}_{i1}R\mathbf{q}_r^0, \tilde{\omega}_{i2}R\mathbf{q}_r^0]
\end{equation}
This process is simple. Note for 2D case, this is even simpler since there is no need for the equation solve in Equ.~\ref{equ:Gomega} since $G^{-1}=tr(S)^{-1}I$.

To compute the Hessian of the shape matching potential energy, we take derivatives over Equ.~\ref{equ:nablav}, where
\begin{equation}
\label{equ:nablaliV}
\nabla^2_{li}V=\sum_rk_r(-(1-\gamma)(\frac{\partial^2R\mathbf{q}_r^0}{\partial\mathbf{q}_l\partial\mathbf{q}_i})^T\mathbf{d}_r+(\frac{\partial\mathbf{d}_r}{\partial\mathbf{q}_i})^T\frac{\partial\mathbf{d}_r}{\partial\mathbf{q}_l})
\end{equation}
To compute the first term of $\nabla_{li}V$, we calculate the derivative of $S$ as
\begin{equation}
\frac{\partial S}{\partial\mathbf{q}_{ls}}=R^T(\frac{m_l}{M}\mathbf{e}_s(\mathbf{q}_l^0)^T-\frac{\partial R}{\partial \mathbf{q}_{ls}}S)
\end{equation}

We take derivatives on both sides of Equ.~\ref{equ:Gomega} and after rearranging (also using the property of antisymmetric matrix where $\tilde{\omega}^T=-\tilde{\omega}$), we have
\begin{equation}
G\omega_{ls,ij}=-\frac{2m_i}{M}\text{skew}(R^T\tilde{\omega}_{ls}\mathbf{e}_j(\mathbf{q}_i^0)^T)-(tr(\frac{\partial S}{\partial\mathbf{q}_{ls}})I-\frac{\partial S}{\partial\mathbf{q}_{ls}})R^T\omega_{ij}+(tr(S)I-S)R^T\tilde{\omega}_{ls}\omega_{ij}
\end{equation}
For 2D case the last term of the right hand side can be dropped since we have $\omega_{ls}\times\omega_{ij}=0$ for any $l,s,i,j$, where we have
\begin{equation}
G\omega_{ls,ij}=-\frac{2m_i}{M}\text{skew}(R^T\tilde{\omega}_{ls}\mathbf{e}_j(\mathbf{q}_i^0)^T)-(tr(\frac{\partial S}{\partial\mathbf{q}_{ls}})I-\frac{\partial S}{\partial\mathbf{q}_{ls}})R^T\omega_{ij}
\end{equation}
After solving for $\omega_{ls,ij}$, we compute $\frac{\partial^2R\mathbf{q}_r^0}{\partial\mathbf{q}_{ls}\partial\mathbf{q}_{ij}}$ as
\begin{equation}
\frac{\partial^2R\mathbf{q}_r^0}{\partial\mathbf{q}_{ls}\partial\mathbf{q}_{ij}}=(\tilde{\omega}_{ls,ij}+\tilde{\omega}_{ij}\tilde{\omega}_{ls})R\mathbf{q}_r^0
\end{equation}
and for 3D,
\begin{equation}
\begin{split}
(\frac{\partial^2R\mathbf{q}_r^0}{\partial\mathbf{q}_l\partial\mathbf{q}_i})^T\mathbf{d}_r=\\
\begin{bmatrix}
{\mathbf{q}_r^0}^TR^T(\tilde{\omega}_{l0}\tilde{\omega}_{i0}-\tilde{\omega}_{l0,i0})\mathbf{d}_r &  {\mathbf{q}_r^0}^TR^T(\tilde{\omega}_{l1}\tilde{\omega}_{i0}-\tilde{\omega}_{l1,i0})\mathbf{d}_r & {\mathbf{q}_r^0}^TR^T(\tilde{\omega}_{l2}\tilde{\omega}_{i0}-\tilde{\omega}_{l2,i0})\mathbf{d}_r \\ 
{\mathbf{q}_r^0}^TR^T(\tilde{\omega}_{l0}\tilde{\omega}_{i1}-\tilde{\omega}_{l0,i1})\mathbf{d}_r &  {\mathbf{q}_r^0}^TR^T(\tilde{\omega}_{l1}\tilde{\omega}_{i1}-\tilde{\omega}_{l1,i1})\mathbf{d}_r & {\mathbf{q}_r^0}^TR^T(\tilde{\omega}_{l2}\tilde{\omega}_{i1}-\tilde{\omega}_{l2,i1})\mathbf{d}_r \\ 
{\mathbf{q}_r^0}^TR^T(\tilde{\omega}_{l0}\tilde{\omega}_{i2}-\tilde{\omega}_{l0,i2})\mathbf{d}_r &  {\mathbf{q}_r^0}^TR^T(\tilde{\omega}_{l1}\tilde{\omega}_{i2}-\tilde{\omega}_{l1,i2})\mathbf{d}_r & {\mathbf{q}_r^0}^TR^T(\tilde{\omega}_{l2}\tilde{\omega}_{i2}-\tilde{\omega}_{l2,i2})\mathbf{d}_r 
\end{bmatrix}
\end{split}
\end{equation}

\section{Shape Matching with Viscous Damping} 
Now we have the equations for undamped motion. Next we derive the force and Hessian of the viscous damping occurred in shape matching. We define
\begin{equation}
\begin{split}
V_{da}=\frac{\alpha}{2}\sum_rk_r\mathbf{\dot{d}}_r^T\mathbf{\dot{d}}_r\\
=\frac{\alpha}{2}\sum_rk_r(\sum_j\frac{\partial\mathbf{d}_r}{\partial\mathbf{q}_j}\mathbf{\dot{q}}_j)^T(\sum_j\frac{\partial\mathbf{d}_r}{\partial\mathbf{q}_j}\mathbf{\dot{q}}_j)
\end{split}
\end{equation}
as the stiffness damping energy, where $\mathbf{\dot{q}}_j$ is the velocity of the $j$-th particle, and
\begin{equation}
V_{db}=\frac{\beta}{2}\sum_rm_r\mathbf{\dot{q}}_r^T\mathbf{\dot{q}}_r
\end{equation}
as the mass damping energy. Following the Rayleigh damping model we have the total damping energy as
\begin{equation}
V_d=V_{da}+V_{db}.
\end{equation}

Hence the potential gradient of the $i$-th particle is
\begin{equation}
\label{equ:nablavd}
\frac{\partial V_d}{\partial \mathbf{\dot{q}}_i}=\beta m_i\mathbf{\dot{q}}_i+\alpha\sum_rk_r(\frac{\partial\mathbf{d}_r}{\partial\mathbf{q}_i})^T\mathbf{\dot{d}}_r
\end{equation}
Correspondingly, the positional Hessian is
\begin{equation}
\label{equ:nablaliVd}
\frac{\partial^2 V_d}{\partial \mathbf{q}_l\partial \mathbf{\dot{q}}_i}=-\alpha(1-\gamma)\sum_rk_r((\frac{\partial^2R\mathbf{q}_r^0}{\partial\mathbf{q}_l\partial\mathbf{q}_i})^T\mathbf{\dot{d}}_r+(\frac{\partial\mathbf{d}_r}{\partial\mathbf{q}_i})^T\sum_j(\frac{\partial^2R\mathbf{q}_r^0}{\partial\mathbf{q}_l\partial\mathbf{q}_j})^T\mathbf{\dot{q}}_j)
\end{equation}
while the velocity Hessian is
\begin{equation}
\frac{\partial^2 V_d}{\partial \mathbf{\dot{q}}_l\partial \mathbf{\dot{q}}_i}=\beta m_i I+\alpha\sum_rk_r(\frac{\partial\mathbf{d}_r}{\partial\mathbf{q}_i})^T\frac{\partial\mathbf{d}_r}{\partial\mathbf{q}_l}
\end{equation}

\section{Total Energy} 
To combine both positional and viscous force, we need to discretize $\mathbf{q}_i$ along time. By denoting $\hat{\mathbf{q}}_i$ as the position of last time step, and $h$ as the time step size, we adopt simple finite difference where
\begin{equation}
\mathbf{\dot{q}}_i=\frac{\mathbf{q}_i-\hat{\mathbf{q}}_i}{h}
\end{equation}
and 
\begin{equation}
\partial\mathbf{q}_i=h\partial\mathbf{\dot{q}}_i
\end{equation}

We then combine Equ.~\ref{equ:nablav} and Equ.~\ref{equ:nablavd}, and we have the Jacobian of shape matching energy as
\begin{equation}
\frac{\partial V_{total}}{\partial \mathbf{q}_i}=\tilde{\beta} m_i\mathbf{\dot{q}}_i+\sum_rk_r(\frac{\partial\mathbf{d}_r}{\partial\mathbf{q}_i})^T(\mathbf{d}_r+\tilde{\alpha}\mathbf{\dot{d}}_r)
\end{equation}
where $\tilde{\alpha}\equiv \alpha h^{-1}$ and $\tilde{\beta}\equiv \beta h^{-1}$.

Similarly, by combining Equ.~\ref{equ:nablaliV} and Equ.~\ref{equ:nablaliVd} we have the positional Hessian of shape matching energy as
\begin{equation}
\frac{\partial^2 V_{total}}{\partial \mathbf{q}_l\partial \mathbf{q}_i}=\sum_rk_r(-(1-\gamma)(\frac{\partial^2R\mathbf{q}_r^0}{\partial\mathbf{q}_l\partial\mathbf{q}_i})^T(\mathbf{d}_r+\tilde{\alpha}\mathbf{\dot{d}}_r)+(\frac{\partial\mathbf{d}_r}{\partial\mathbf{q}_i})^T(\frac{\partial\mathbf{d}_r}{\partial\mathbf{q}_l}-(1-\gamma)\tilde{\alpha}\sum_j(\frac{\partial^2R\mathbf{q}_r^0}{\partial\mathbf{q}_l\partial\mathbf{q}_j})^T\mathbf{\dot{q}}_j))
\end{equation}

{
\bibliographystyle{alpha} 
\bibliography{ref}
}
\end{document}